\documentclass[a4paper,11pt]{article}
\usepackage{amsmath}
\usepackage{amssymb}
\usepackage{graphicx}
\usepackage{cite}
\usepackage[T1]{fontenc}
\usepackage[a4paper,left=1.0in, right=1.0in,top=1.1in, bottom=1.2in]{geometry}

\newcommand{\bit}{\begin{itemize}}
\newcommand{\eit}{\end{itemize}}
\newcommand{\be}{\begin{equation}}
\newcommand{\ee}{\end{equation}}
\newcommand{\bea}{\begin{eqnarray}}
\newcommand{\eea}{\end{eqnarray}}
\newcommand{\bsp}{\begin{split}}
\newcommand{\esp}{\end{split}}

\title{ 
    \vskip 2cm
    Kinematical constraint effects in the evolution equations based on angular ordering 
}

\author{
  Michal Deak\\
{\it Instituto de F\'isica Corpuscular,}\\ {\it Universitat de Val\`encia - Consejo Superior de Investigaciones Cient\'ificas,}\\
{\it Parc Cient\'ific, E-46071 Paterna (Val\`encia), Spain}\\
\\
Krzysztof Kutak\\
{\it Instytut Fizyki Jadrowej, Polskiej Akademii Nauk,}\\
{\it Radzikowskiego 152, 31-342 Krakow, Poland}
  }
\date{}

\begin{document}
\maketitle
\vspace{-25em}
\begin{flushright}
  IFJPAN-IV-2015-2
\end{flushright}
\vspace{20em}

\begin{abstract}
We study effects of imposing various forms of the kinematical constraints on the full form of the CCFM equation and its non-linear extension. We find, that imposing the constraint in its complete form modifies significantly the shape of gluon density as compared to forms of the constraint used in numerical calculations and phenomenological applications. In particular the resulting gluon density is suppressed for large values of hard scale related parameter and $k_T$ of gluon. This result might be important in description of jet correlations at Large Hadron Collider within CCFM approach.
\end{abstract}

\section*{Introduction}
In the high energy limit of hadron scattering, in the regime where the center 
of mass energy is larger than any other available scale, perturbative approach 
to processes with high momentum transfer allows factorization 
of the cross section into a hard matrix element with initial off-shell gluons 
and an unintegrated gluon 
density~\cite{Gribov:1984tu,Catani:1990eg}. The unintegrated gluon density is a 
function of the 
longitudinal momentum fraction $x$ and transverse momentum ${k_T}$ of a gluon.
After taking into account formally subleading corrections coming from coherence 
of gluon emissions, one is lead to the CCFM 
set of equations~\cite{Ciafaloni:1987ur,Catani:1989sg,Catani:1989yc} which 
introduce gluon density dependent on hard scale related to the probe as well as one
introduces unintegrated quark densities.
The status of phenomenological relevance of CCFM framework is not fully satisfactory. In 
principle, it is a set of equations that should be the ideal framework for application 
to final states at high energies and covering DGLAP and BFKL domains. It has been implemented in the Monte Carlo event generator \cite{Jung:2010si}.
However, so far good agreement with high precision data has been successfully achieved only in rather inclusive processes like  $F_2$ and Drell-Yan \cite{Hautmann:2013tba}. It is known that on the theory side the CCFM physics is still to be completed. Below we outline the main points. For 
instance:
\begin{itemize}
\item
the CCFM system of equations has been so far solved in decoupled approximation in the, gluonic (for short the equation for gluon density is usually called CCFM) and non-singlet 
channels neglecting correlations between partons \cite{Hautmann:2014uua}.
This might result 
in improper treatment of the kinematical region where the gluon degrees of freedom are less 
dominant than sea quarks. Since gluon might be artificially dominating over 
quarks where it should actually be suppressed. Furthermore the corrections coming from inclusion of transversal momentum dependence to splitting functions $P_{gq}$ and $P_{qq}$ are not known and in order to have complete picture they should be calculated \cite{Hautmann:2014uua}.
\item 
the impact of kinematical effects introducing energy conservation in the CCFM evolution 
have not been investigated in all detail. As it turns out from our study it is necessary to 
revisit the inclusion of the so called kinematical constraint~\cite{Kwiecinski:1996td} into the CCFM equation. 
The orgin of the kinematical constraint follows from refining the assumption that the $t$-channel off-shell gluon's $4$-momentum is 
dominated by $k_T^2$. The kinematical constraint causes suppression of the
gluon density and even overrides the angular ordering in regions of large ${k_T}$.
\item only recently the CCFM has been promoted to non-linear 
equation\cite{Kutak:2011fu,Kutak:2012yr,Kutak:2012qk} therefore allowing 
for the possibility to investigate interplay of coherence effects and 
saturation \cite{Gribov:1984tu} in exclusive processes like dijet production at the LHC 
~\cite{Deak:2009xt,Deak:2009ae,Deak:2010gk,Kutak:2012rf,Chatrchyan:2012gwa,vanHameren:2014lna,vanHameren:2014ala,Kutak:2014wga}.
In particular the important question is what is the role of the angular ordering 
and kinematical effect in the evolution at the non-linear level. The optimal 
form of initial conditions is also not known.   
\end{itemize}
This publication is a continuation of the work done in~\cite{Deak:2012mx,Kutak:2013yga}. We compare numerically forms of non-Sudakov form factor used in the literature and solve the full CCFM equation and its non-linear extension -- including the 
$1/(1-z)$ pole and kinematical constraint in the kernel of the equation i.e. not only in the non-Sudakov form factor. However we keep  $\alpha_s$ 
constant in order to have clear picture of the role of the kinematical effects. \\
\begin{figure}[t!]
\label{ref:fig0}
\begin{picture}(30,30)
   \put(50, -240){
      \includegraphics{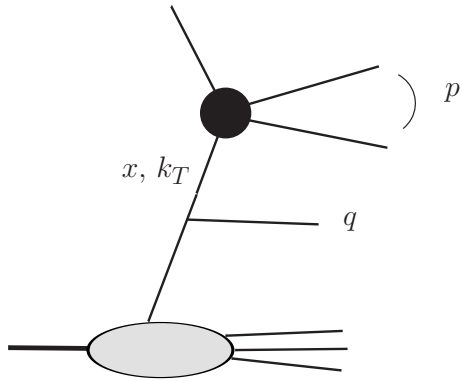}
    }

      \end{picture}
\vspace{2cm}
\caption{\small The plot visualizing kinematical variables used in the CCFM equation. The black blob represents hard process while the gray oval represents hadron.}
\vspace{0.5cm}
\label{fig:fig1}
\end{figure}
\section*{The CCFM equation}
\label{sec:ccfm}
The CCFM equation for gluon density reads:
\begin{equation}\label{eq:IS-CCFM}
\begin{split}
\mathcal{A}(x,{k_T},p)&=\mathcal{A}_0(x,{k_T},p)+{\bar\alpha}_S
\int\frac{d^2\bar{\bf q}}{\pi\bar{\bf q}^{2}}
\int\limits_{x}^{1-\frac{Q_0}{\bar q}}dz\;
\;\theta(p-z{\bar{q}})\;{\mathcal P}(z,{k_T},\bar{q})\;
\Delta_S(p,z\bar{q},Q_0)\; \mathcal{A}(x/z,{k_T}^{\prime},\bar{q}),
\end{split}
\end{equation}
where $k^\prime=|{\bf k}+(1-z){\bf \bar q}|$ and the moduli of the two dimensional vectors transversal to the collision plane are denoted $|{\bf k}|\equiv k_T$, $|{\bf q}|\equiv q_T$, $x$ is gluon's longitudinal momentum fraction and $\bar\alpha_s=N_c\alpha_s/\pi$.
Also the rescaled momentum is introduced as $\bar{q}=|\bar{\bf q}|={q_T}/(1-z)$.\\
The splitting function assumes the form:
\begin{eqnarray}
{\mathcal P}\left(z,{k_T},\bar q\right)  &=&
\frac{\Delta_{NS}(z,k_T^2,\bar{q})}{z}+\frac{1}{1-z}\;.
\end{eqnarray}

The Sudakov form-factor evaluated in double logarithmic approximation reads:
\begin{equation}
\label{eq:sud}
\Delta_S(p,z\bar{q},Q_0)=\exp
\Bigg(-\int\limits_{(z\bar{\bf q})^2}^{p^2}
\frac{d{q^{'2}}}{{{q}^{\prime}}^2}
\int\limits_{0}^{1-\frac{Q_0}{|{\bf q}^\prime|}}dz^\prime
\frac{\bar\alpha_S}{1-z^\prime}\Bigg)\;.
\end{equation}

The non-Sudakov form-factor regularizing $1/z$ singularity in angular ordered 
region is:
\begin{equation}
\label{eq:nonsud1}
\Delta_{NS}(z,k_T,\bar{q})=\exp\Bigg(-\bar\alpha_s\int_z^1
\frac{dz^\prime}{z^\prime}\int_{z^2{\bar q}^2}^{k^2}
\frac{d^2q^\prime}{q^{2\prime}}\Bigg)=\exp\left(-\bar\alpha_S\ln\frac{1}{z}
\ln\frac{k_T^2}{z{\bar q}^2}\right)\; .
\end{equation}
\subsection*{The kinematical constraint}\label{sec:kc}
The integration over $\bar{\bf q}$ in the equation~\eqref{eq:IS-CCFM},
although being constrained from below by the soft cut-off $Q_0$, is not 
constrained by an upper limit thus violating the energy-momentum conservation. 
Moreover in the low $x$ formalism one requires that in the denominator of the 
off-shell gluon propagator one keeps terms that obey $|k^2|=k_T^2$.
In order to be consistent as argued in \cite{Catani:1989sg} the non-Sudakov form-factor 
should be accompanied by a kinematical constraint limiting the above 
integration over $\bar q$. In approximated form it reads
\vspace{0.2cm}
\begin{equation}
\label{eq:kincon}
{k_T^2}>z\,\bar q^2
\end{equation}
After including it in the (\ref{eq:nonsud1}) we have:
\begin{equation}
\label{eq:nonsud10}
\Delta_{NS}(z,k_T,\bar{q})=\exp\left(-\bar\alpha_S\ln\frac{1}{z}\ln\frac{k_T^2}{z{\bar q}^2}\theta(k_T^2-z\bar q^2)\right)\; .
\end{equation}
The condition~\eqref{eq:kincon} at $z\ll 1$ guaranties that $|k^2|\simeq k_T^2$.
In \cite{Kwiecinski:1996td} it has been extended to region including also the case when 
$z\sim 1$. Below we re-obtain the full form of kinematical constraint emphasizing 
the its role in conservation of energy.
Having $k=z\,p^++{\bar z}p^-+k_{\perp}$ with $p^+$ and $p^-$ being the
initial state gluon momenta $+$ and $-$ components,
we can write the expression for $k^2$
\begin{equation}\label{eq:cutoff}
k^2=-z\,{\bar z}\,{\hat s}-k_T^2\;,
\end{equation}
with ${\hat s}=(p^++p^-)^2=2p^+\cdot p^-$. 
Note, that for the full propagator $|k^2|>k_T^2$, thus the full propagator 
causes stronger suppression of the amplitude which has to be taken into account 
when using $|k^2|\simeq k_T^2$. A cut-off like~\eqref{eq:cutoff} is a simple 
implementation of the suppression.\\
The condition $|k^2|\simeq k_T^2$ translates approximately to
\begin{equation}\label{eq:ksqs0app}
k_T^2>z\,{\bar z}\,{\hat s}.
\end{equation}
Using the identity $q^2={\bar z}(1-z){\hat s}-{q_T}^2=0$, we can express 
${\bar z}\,{\hat s}={q_T}^2/(1-z)$
and insert it into~\eqref{eq:ksqs0app} to obtain:
\begin{equation}\label{eq:kincon-f}
k_T^2>\frac{z\,{q_T}^2}{1-z}=\,z\,(1-z)\,{\bar q}^2
\end{equation}
\begin{figure}[t!]
\label{ref:fig1}
\begin{picture}(30,30)
   \put(-3, -110){
      \includegraphics{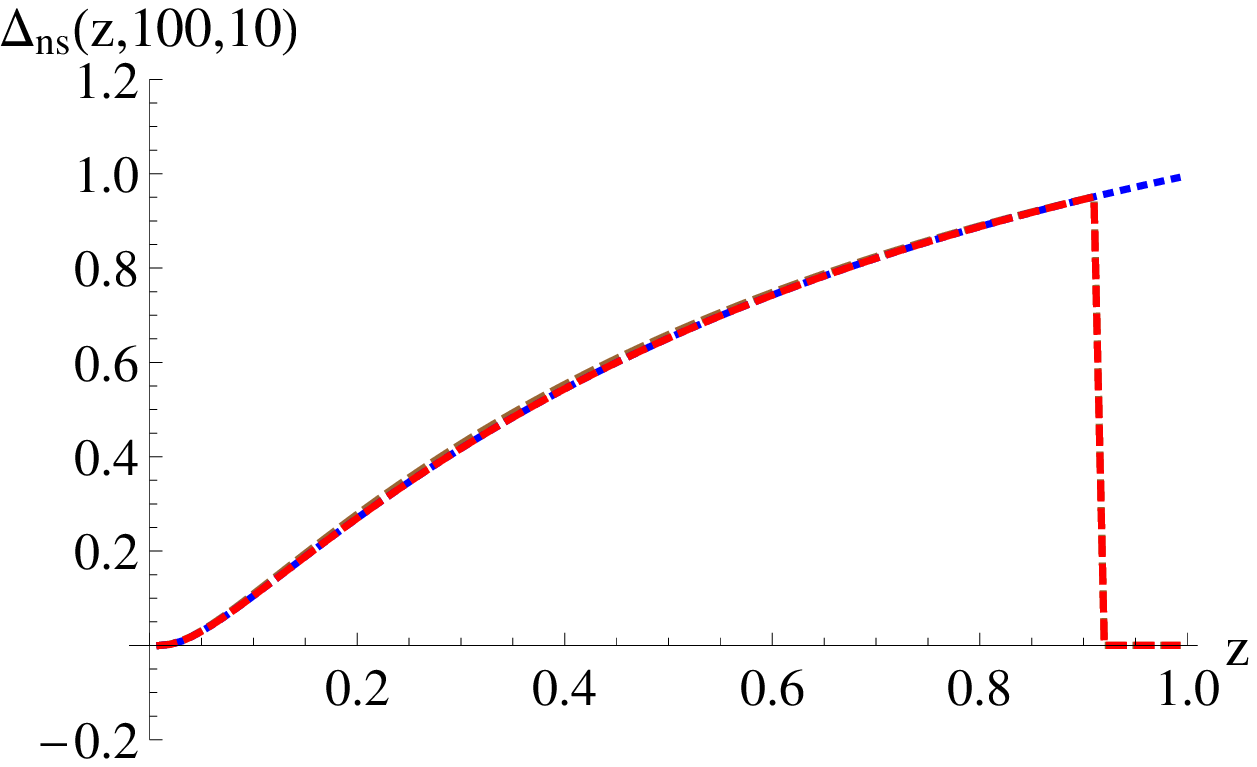}
    }

\put(222, -110){
      \includegraphics{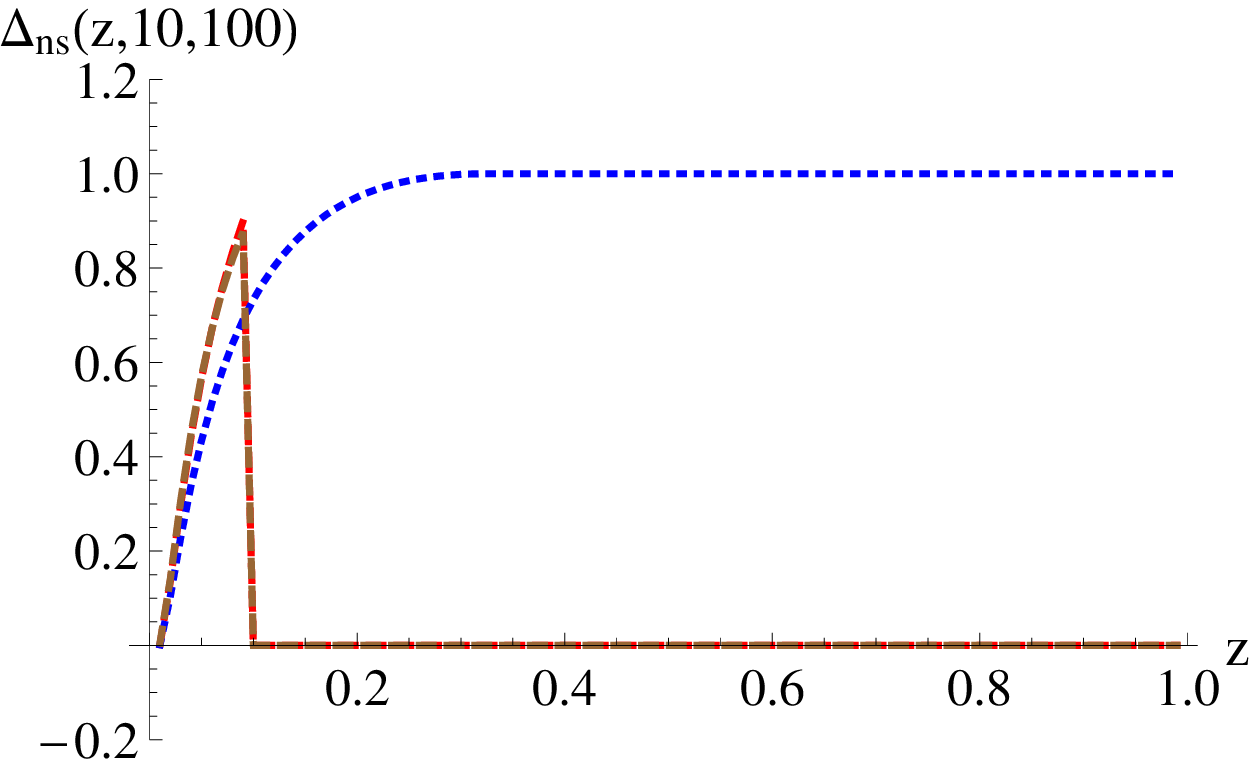}
    }

\end{picture}
\vspace{4cm}
\caption{\small The plots visualizes the various forms of non-Sudakov form 
factor. On the left we choose the situation when $k_T=10GeV$ and 
$\bar q=4 GeV$ while on the right $k_T=4GeV$ and $\bar q=10 GeV$. The red
line corresponds to form-factor given by formula \eqref{eq:nonsud10}, the brown 
line corresponds to form-factor given by formula \eqref{eq:nonsud2} while blue 
line to form-factor given by formula \eqref{eq:nonsud11}. The form-factor given 
by \eqref{eq:nonsud2} is multiplied by the function 
$\theta\left(k_T/\left(1-z\right){\bar q}^2-z\right)$}.
\vspace{0.5cm}
\label{fig:fig1}
\end{figure}
The lower bound on $z>x$ results in the upper bound on 
$q_T^2<k_T^2/x\simeq{\hat s}$ providing local condition for energy-momentum 
conservation.
The CCFM with the kinematical constraint included reads \cite{Kwiecinski:1996td}:
\begin{equation}\label{eq:ccfm-or}
\begin{split}
\mathcal{A}(x,{k_T},p)&=\mathcal{A}_0(x,{k_T},p)+{\bar\alpha}_S
\int\frac{d^2\bar{\bf q}}{\pi\bar{\bf q}^{2}}
\int\limits_{x}^{1-\frac{Q_0}{|{\bar{\bf q}}|}}dz\;
\theta\left(\frac{k_T^2}{(1-z)\bar{q}^2}-z\right)\\
&\times\;\theta(p-z|{\bar{q}}|)\;{\mathcal P}(z,{k_T},\bar{q})\;
\Delta_S(p,z\bar{q},Q_0)\mathcal{A}(x/z,{k_T}^{\prime},\bar{q})\;,
\end{split}
\end{equation}
and the non-Sudakov form-factor after inclusion of the full form of the 
kinematical constraint assumes form
\begin{equation}
\label{eq:nonsud2}
\begin{split}
\Delta_{NS} (z,k_T, \bar{q})= \exp \bigg\{ -
\overline{\alpha}_S & \int_z^1 \frac{dz^\prime}{z^\prime} \; \Theta
\left ( \frac{(1-z^\prime)k_T^2}{(1-z)^2\bar q^2} - z^\prime \right )\times\\ \: &\int \:
\frac{d{q}^{\prime 2}}{{q}^{\prime 2}} \; \Theta (k_T^2 -
{q}^{\prime 2}) \; \Theta ({q}^{\prime} - z^\prime\bar q) \bigg\}\;.
\end{split}
\end{equation}
Please note the presence of the function $\theta\left(\frac{k_T^2}{(1-z)\bar{q}^2}-z\right)$. 
At the level of the kernel of the BFKL equation obtained directly from Feynmann diagrams (virtual and real corrections are kept apart) one imposes the kinematical constraint at the so called "unresummed" level on real emissions. 
However in \cite{Kwiecinski:1996td} it has been observed, at a level of BFKL equation, that imposing the kinematical constraint on unresummed equation and performing algebraic transformations leading to resummed equation one obtains a $\theta$-function affecting the full kernel. The starting point of the CCFM is a resummed kernel, therefore this result suggested how the complete form of the 
kinematical constraint should be imposed on the kernel of CCFM. The authors of \cite{Kwiecinski:1996td} solve the CCFM at small $z$ limit therefore the function $\theta\left(\frac{k_T^2}{(1-z)\bar{q}^2}-z\right)$ is neglected and in most of the  phenomenological and theoretical applications of the CCFM this term is neglected 
\cite{Deak:2010gk,Hautmann:2013tba,Salam:1998cp,Bottazzi:1998rs,Salam:1999ft,Avsar:2010ia,Bacchetta:2010hh,Chachamis:2011rw} as well.
The following form of non-Sudakov form factor is usually being used:
\begin{eqnarray}
\label{eq:nonsud11}
\Delta_{NS}(z,k_T,\bar{q}) & = & \exp \left ( - {\overline \alpha}_S
\int_z^{z_0} \frac{dz^\prime}{z^\prime} \int \frac{dq^{\prime 2}}{q^{\prime 2}}
\Theta
({k}^2 - q^{\prime 2}) \Theta (q^{\prime} - z^\prime \bar q) \right ) \label{a8}\\
& = & \exp \left ( - {\overline \alpha}_S \log \left (
\frac{z_0}{z} \right ) \log \left ( \frac{{k}^2}{z_0z{\bar q}^2} \right
) \right )\nonumber,
\end{eqnarray}
\noindent where
\begin{equation}
z_0 = \left \{ \begin{array}{lll}
1\text{    ,} & {\rm if} & ({k_T}/{\bar q}) \geq 1 \\
{k_T}/{\bar q}\text{,} & {\rm if} & z < ({k_T}/{\bar q}) < 1 \\
z\text{    ,} & {\rm if} & ({k_T}/{\bar q}) \leq z
\end{array} \right.
\end{equation}
and the discussed above the $\theta$-function is not taken into account.\\ As shown in the Fig.~\ref{fig:fig1} the simplified non-Sudakov form-factor \eqref{eq:nonsud10}
approximates well the exact one \eqref{eq:nonsud2} in regions of $z$ where the $\theta$-function is not affecting the kernel while the (\ref{eq:nonsud11}) is quite different. 
In our analysis we consider following equations:
\begin{equation}\label{eq:ccfm-or1}
\begin{split}
\mathcal{A}(x,{k_T},p)&=\mathcal{A}_0(x,{k_T},p)+{\bar\alpha}_S
\int\frac{d^2\bar{\bf q}}{\pi\bar{\bf q}^{2}}
\int\limits_{x}^{1-\frac{Q_0}{\bar q}}dz\;
\theta\left(\frac{k_T^2}{(1-z)\bar{q}^2}-z\right)\\
&\times\;\theta(p-z{\bar q})\;{\mathcal P}(z,{k_T},\bar{q})\;
\Delta_S(p,z\bar{q},Q_0)\mathcal{A}(x/z,{k_T}^{\prime},\bar{q})\;,
\end{split}
\end{equation}
where the splitting function includes non-Sudakov given by formula \eqref{eq:nonsud2} and
\begin{equation}\label{eq:ccfm-or2}
\begin{split}
\mathcal{A}(x,{k_T},p)&=\mathcal{A}_0(x,{k_T},p)+{\bar\alpha}_S
\int\frac{d^2\bar{\bf q}}{\pi\bar{\bf q}^{2}}
\int\limits_{x}^{1-\frac{Q_0}{\bar q}}dz\;\\
&\times\;\theta(p-z{\bar{q}})\;{\mathcal P}(z,{k_T},\bar{q})\;
\Delta_S(p,z\bar{q},Q_0)\mathcal{A}(x/z,{k_T}^{\prime},\bar{q})\;,
\end{split}
\end{equation}

where the non-Sudakov given by formula \eqref{eq:nonsud11}.

\subsection*{Saturation effects and kinematical constraint combined}
To account for gluon recombination at large gluon densities the CCFM equation 
has been promoted to non-linear equation by including a quadratic term \cite{Kutak:2011fu,Kutak:2012yr,Kutak:2012qk} which
reads:
\begin{equation}\label{eq:IS-KGBJS}
\begin{split}
\mathcal{A}(x,{k_T},p)&=\mathcal{A}_0(x,{k_T},p)+{\bar\alpha}_S
\int\frac{d^2\bar{\bf q}}{\pi\bar{\bf q}^{2}}
\int\limits_{x}^{1-\frac{Q_0}{\bar q}}dz\;\theta\left
(\frac{k_T^2}{(1-z)\bar{q}^2}-z\right)\theta(p-z{\bar{q}})\;{\mathcal P}(z,{k_T},{\bar q})\\
&\times\;\Delta_S(p,z\bar{q},Q_0)\;\left(\mathcal{A}(x/z,{k_T}^{\prime},\bar{q})-
\delta\left(\bar{q}^2-\frac{{
k_T}^2}{(1-z)^2}\right){\bar{q}}^2\;\mathcal{A}^2(x/z,{\bar{q}},\bar{q})\right)
\;,
\end{split}
\end{equation}
where we also included the kinematical constraint of the form \eqref{eq:nonsud2} 
in the kernel.
Simpler versions of the equation above have been already analyzed in~\cite{Kutak:2013yga} 
and it has been observed that 
\begin{itemize}
\item the equation leads to phenomenon called saturation at the saturation scale \cite{Avsar:2010ia,Kutak:2013yga}
\item the saturation strongly suppresses the gluon density at low $x$ and low
$k_T$
\end{itemize}
The natural question arises how are these results modified when some of the 
approximations are not taken and how are they modified  if the 
kinematical effect is imposed in the full form.
\begin{figure}[t!]
\vspace{2.5cm}
  \begin{picture}(30,0)
    \put(10, -105){
      \includegraphics{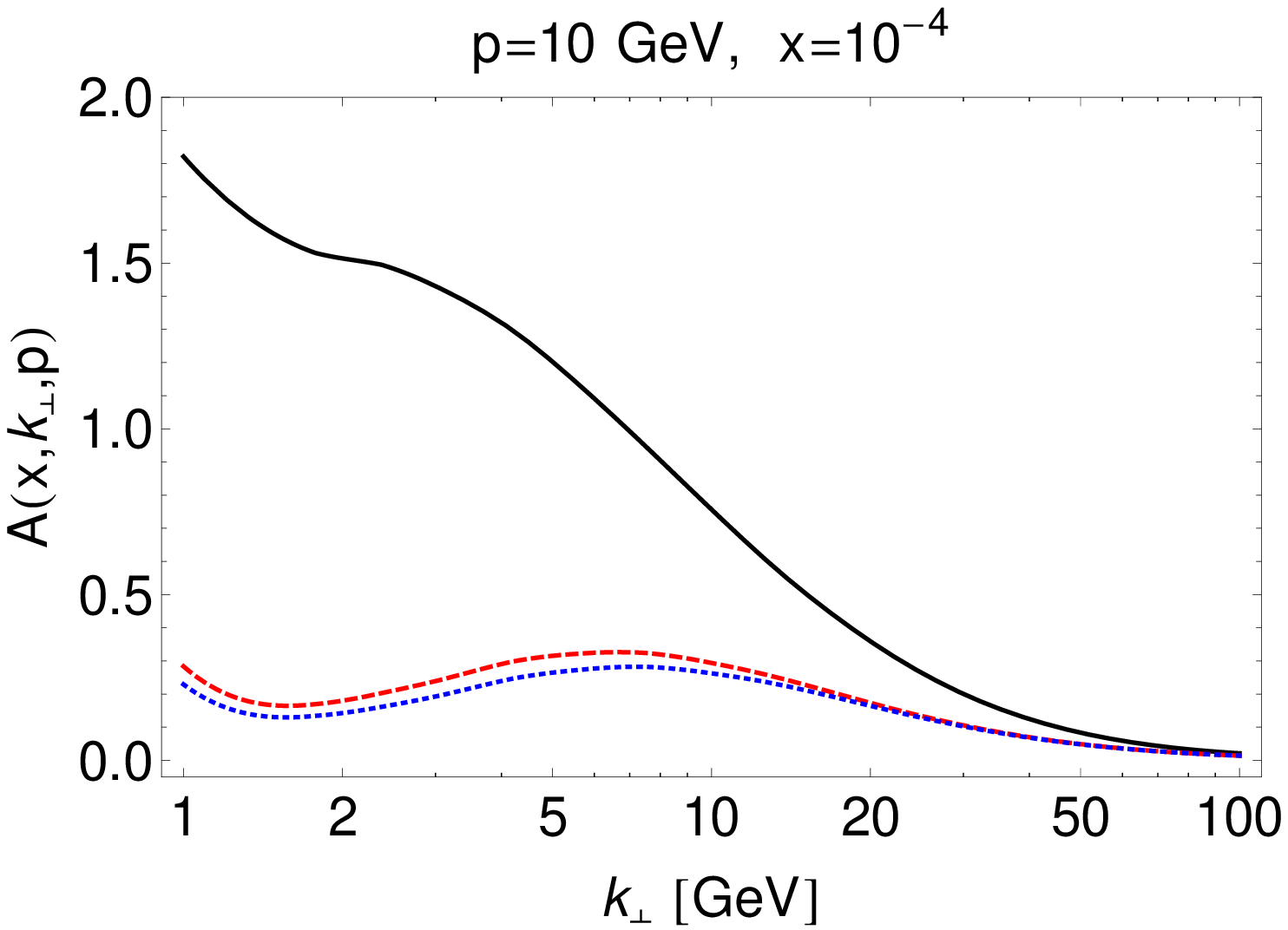}
    }
    \put(230, -107.5){
      \includegraphics{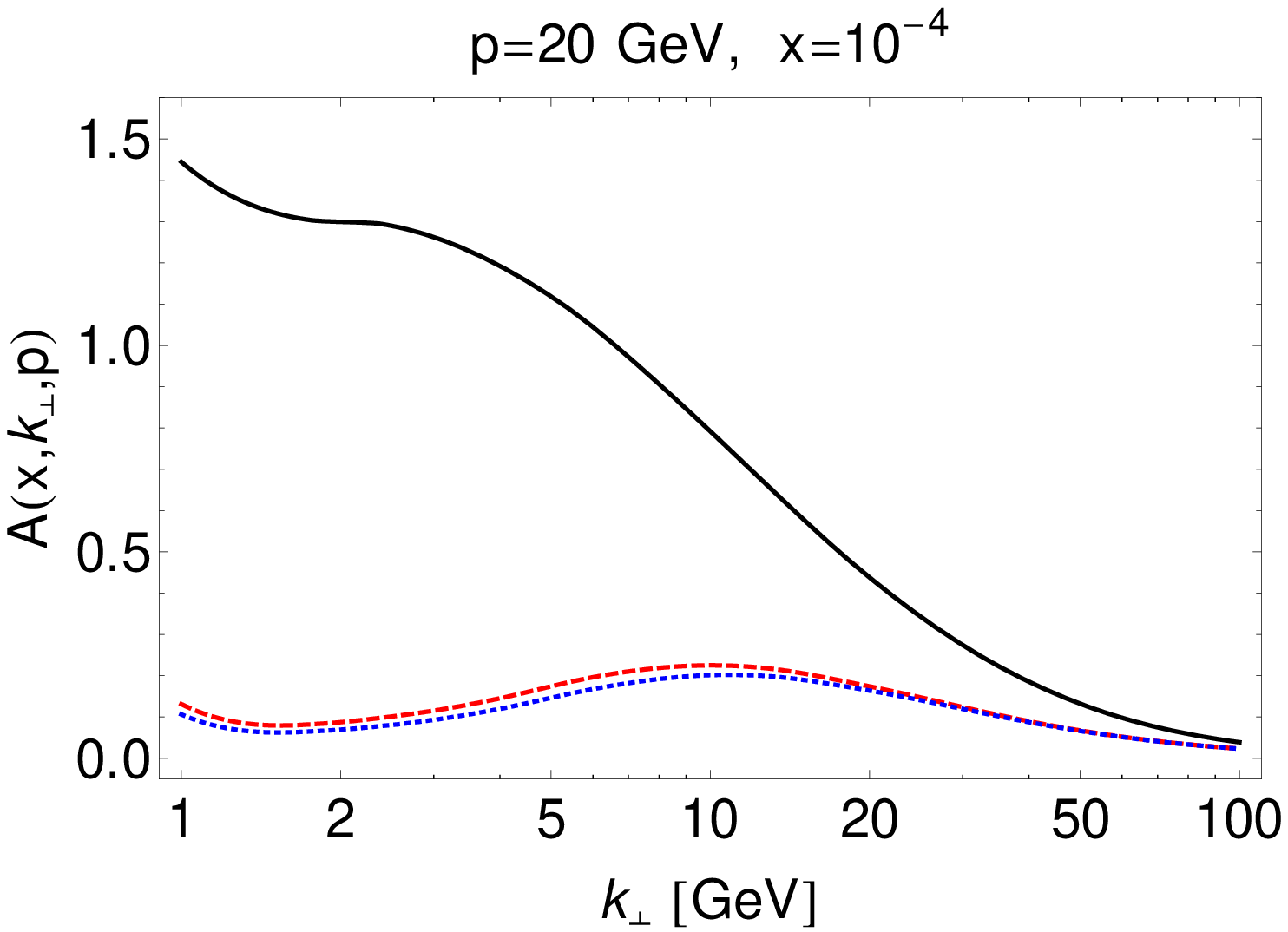}
      }
          
     \end{picture}
\vspace{3.6cm}
\caption{Red (dashed) - equation~\eqref{eq:ccfm-or1}, 
black (solid) - equation~\eqref{eq:ccfm-or2}, blue (dotted) - equation~\eqref{eq:IS-KGBJS}.}
\label{fig:plots1D-1}
\end{figure}
\begin{figure}[tbh]
\vspace{2.5cm}
  \begin{picture}(30,0)
    \put(10, -105){
      \includegraphics{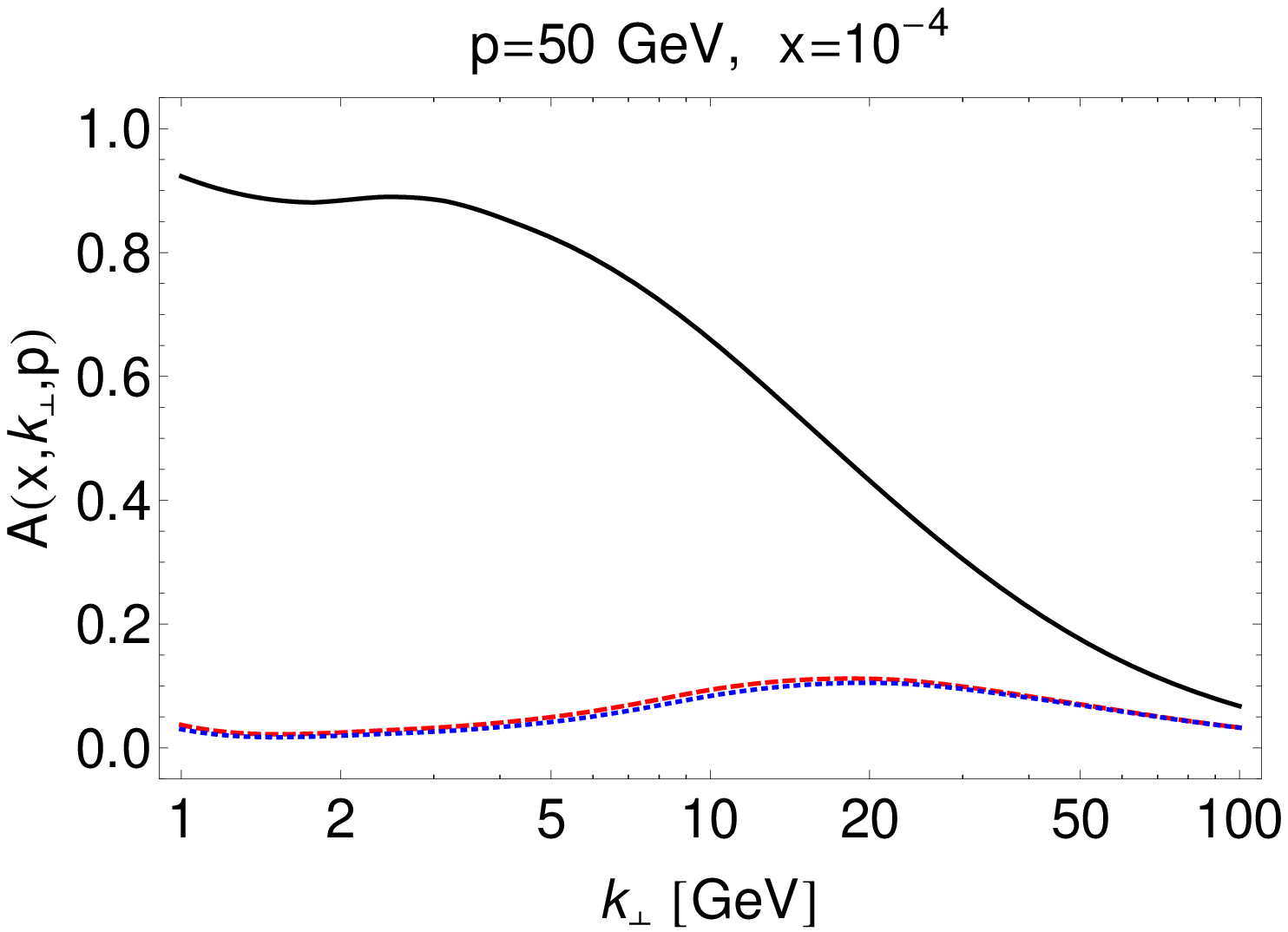}
    }    
    \put(230, -108){
      \includegraphics{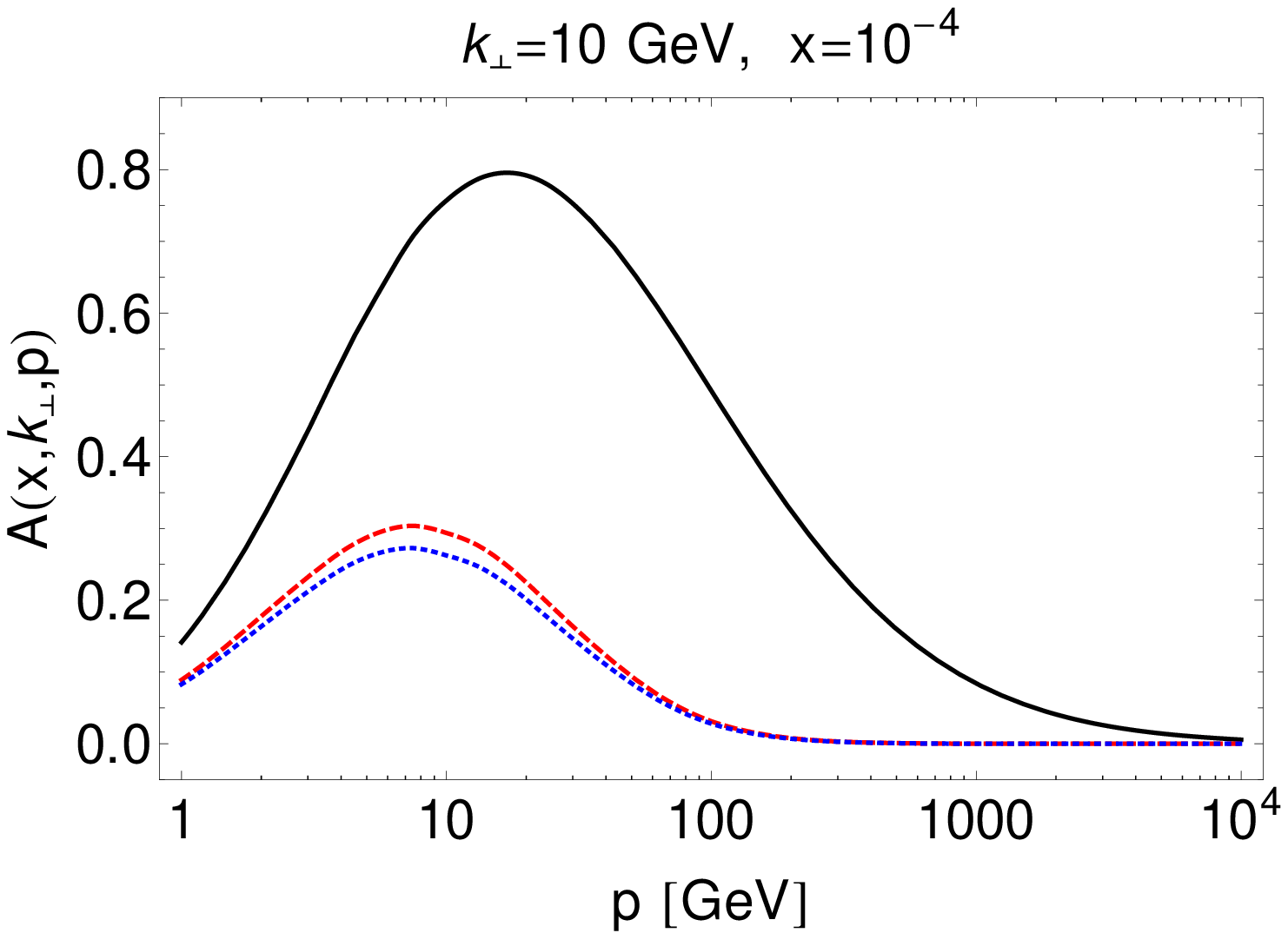}
    }
          
     \end{picture}
\vspace{3.6cm}
\caption{Red (dashed) - equation~\eqref{eq:ccfm-or1}, 
black (solid) - equation~\eqref{eq:ccfm-or2}, blue (dotted) - equation~\eqref{eq:IS-KGBJS}.}
\label{fig:plots1D-2}
\end{figure}
\begin{figure}[tbh]
\vspace{2.5cm}
  \begin{picture}(30,0)      
    \put(10, -105){
      \includegraphics{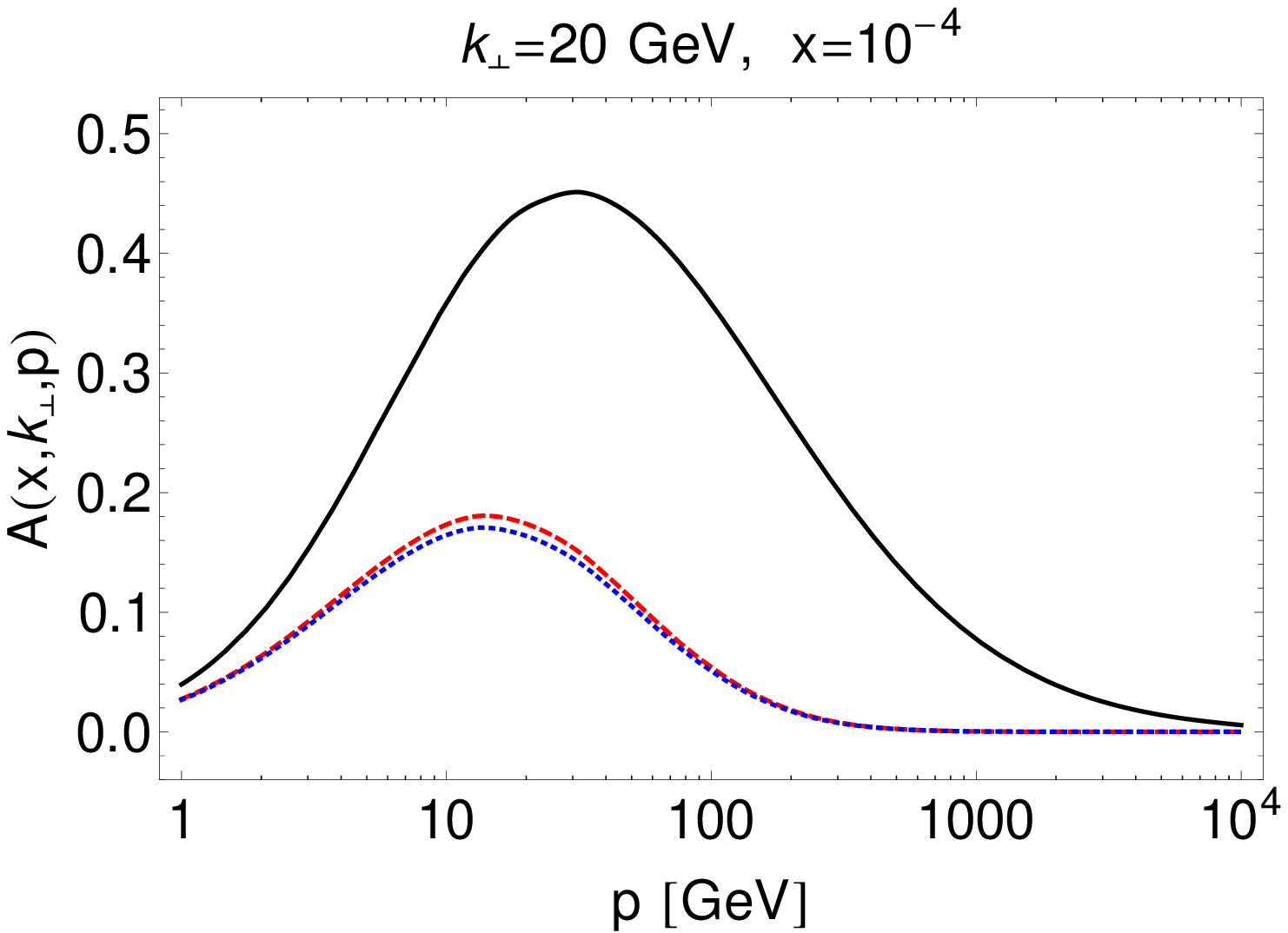}
    }    
    \put(230, -107){
      \includegraphics{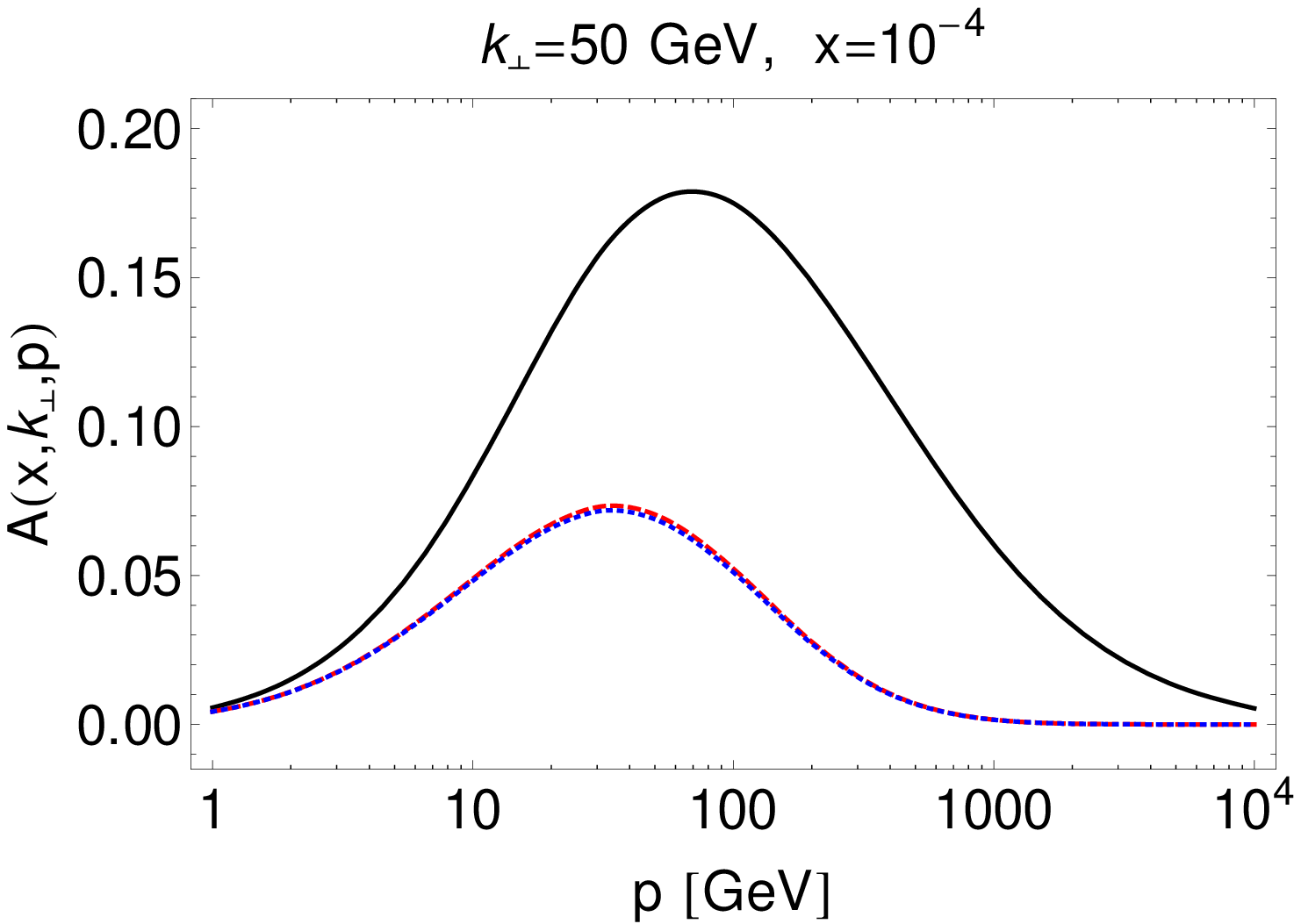}
    }
          
     \end{picture}
\vspace{3.6cm}
\caption{Red (dashed) - equation~\eqref{eq:ccfm-or1}, 
black (solid) - equation~\eqref{eq:ccfm-or2}, blue (dotted) - equation~\eqref{eq:IS-KGBJS}.}
\label{fig:plots1D-3}
\end{figure}
\begin{figure}[tbh]
\vspace{2.5cm}
  \begin{picture}(30,0)      
    \put(10, -105){
      \includegraphics{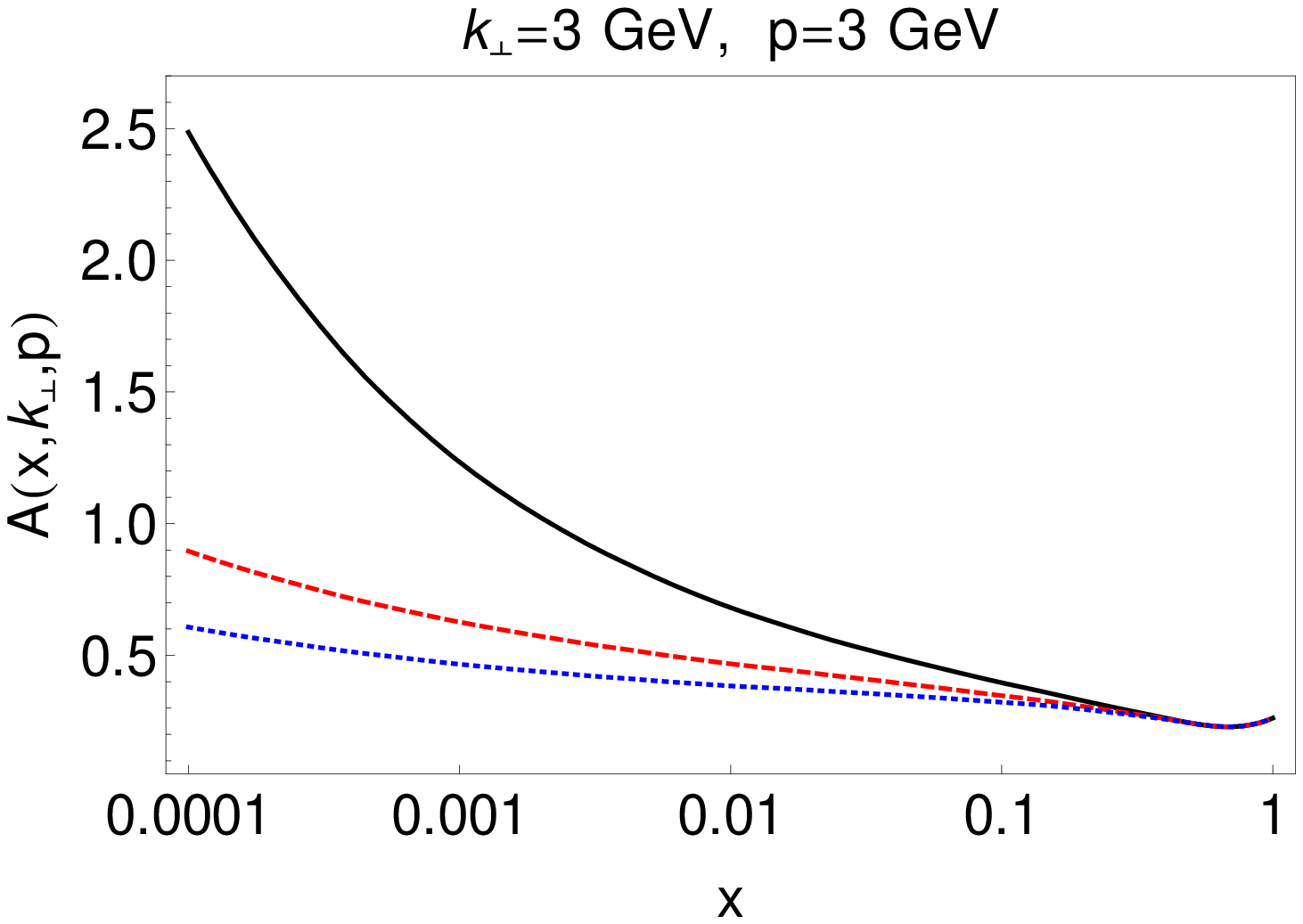}
    }    
    \put(230, -107){
      \includegraphics{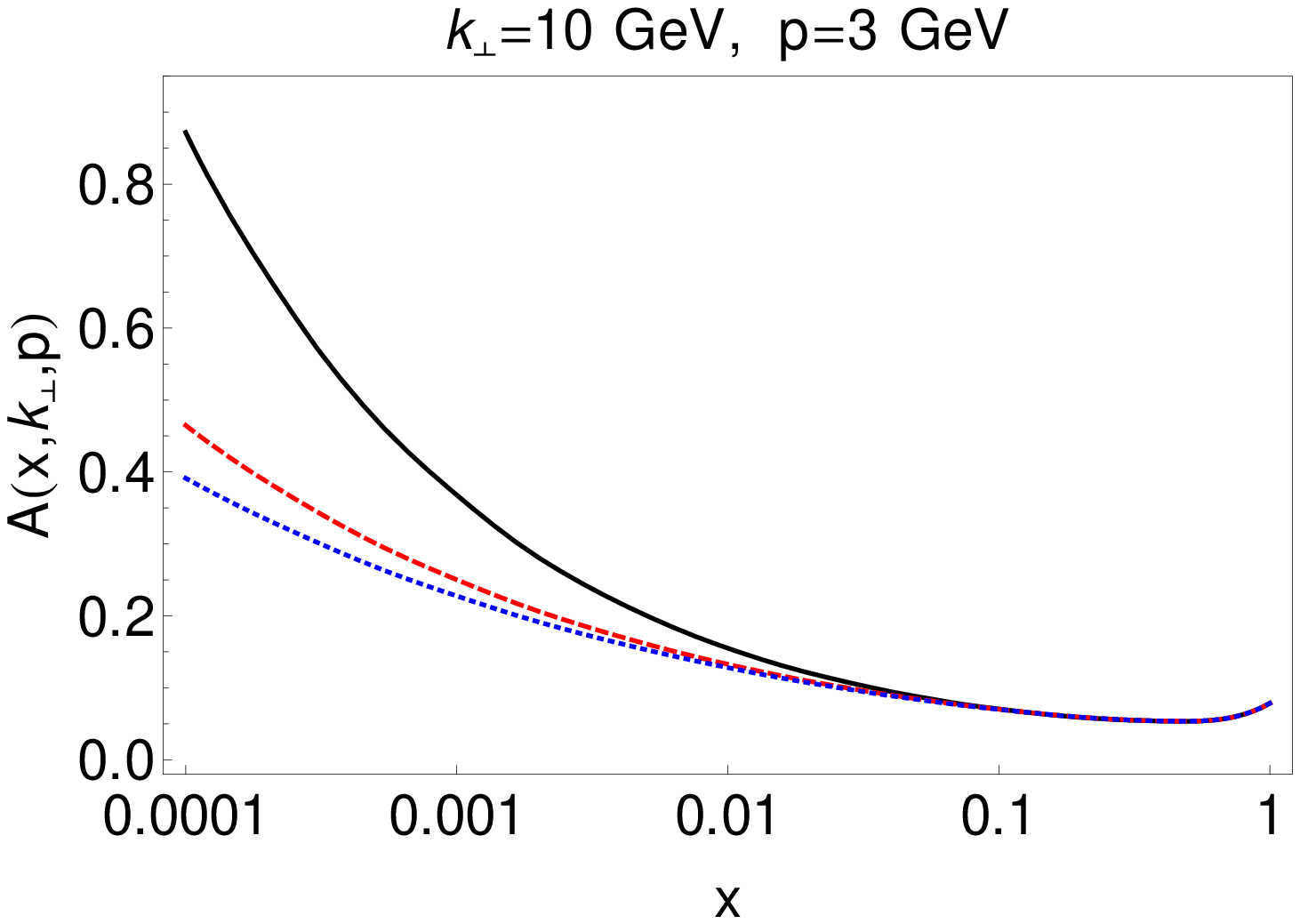}
    }
          
     \end{picture}
\vspace{3.6cm}
\caption{Red (dashed) - equation~\eqref{eq:ccfm-or1}, 
black (solid) - equation~\eqref{eq:ccfm-or2}, blue (dotted) - equation~\eqref{eq:IS-KGBJS}.}
\label{fig:plots1D-4}
\end{figure}



\section*{Numerical results and the discussion}\label{sec:results}
We present the numerical results with ${\bar\alpha}_S=0.2$ and the parameter $Q_0$ set to $Q_0=1\;$GeV. To
study the variation of the results depending on $Q_0$ we produce a solution also for
$Q_0=0.5\;$GeV. The choices of the starting scale are motivated by a
possibility to 
compare the equations in a region where both linear and non-linear equations are phenomenologically relevant.

\subsection*{Linear equations}
We use 
an initial condition which includes resummed virtual and unresolved contributions, 
according to~\cite{Kutak:2013yga} and~\cite{Bacchetta:2010hh}, in the form
\begin{equation}
{\mathcal
A}_0\left(x,k_T,p\right)=A\,\Delta_R(x,k_T,Q_0)\,\Delta_S(p,Q_0,Q_0)/k_T,
\end{equation}
where $A=1/2$ and 
$\Delta_R(z,k_T,Q_0)=\exp\left(-{\bar\alpha}_S\log\frac{1}{z}\log\frac{k^2_T}{Q^2_0}\right)$
is the Regge form-factor.

The first observation we make from the plots in 
Fig.~\ref{fig:plots1D-1}--\ref{fig:plots1D-3} is that the solutions of
equations we study differ significantly. The solutions exhibit also similar features. 
Solutions of both versions of the kernel with kinematical constraint exhibit a local 
maximum as functions of $k_T$ and $p$ with $x$ and $p$ or $k_T$ correspondingly fixed. 
The positions of local maximum in the plots of $p$ dependence are correlated with the value
of $k_T$, with a shift to higher $k_T$ for the solution of the equation~\eqref{eq:ccfm-or2}. 
The peak can be explained by the fact that 
the contribution of the integral on the right hand side of~\eqref{eq:ccfm-or1} peaks at 
around $k_T\sim p$. To point out: the peak is a result of presence of
$\theta\left(p-z{\bar q}\right)$ -- angular ordering condition.
Similar peaks are present also in the plots of $k_T$ dependence and resemble Sudakov suppression of $k_T$ scales of the order of $p$ in case of equation~\eqref{eq:ccfm-or1} \cite{Kutak:2014wga}. 
However, in the 
case of~\eqref{eq:ccfm-or2} it seems, that the position of the peak does not depend on the
value of $p$.
It seems that the kernel with the kinematical constraint included in an implicit
way~\eqref{eq:ccfm-or2} produces similar effect, though weaker, as an implicit inclusion of 
the $\theta$-function only in the case of $p$ dependence, but for the $k_T$ dependence
leads to much smaller suppression. Thus the peak observed in solution
of~\eqref{eq:ccfm-or2} is 'buried' under the result of the evolution. We can conclude that
the peak in the $k_T$ dependence is a result of an interplay of the inclusion of the explicit 
$\theta\left(\frac{k_T^2}{(1-z)\bar{q}^2}-z\right)$ factor and the Sudakov effect.

\subsection*{Non-linear equation}
We set the parameter characterizing the strength of the non-linear term $R$ the value  $R=\sqrt{1/\pi}\;$GeV in the 
equation~\eqref{eq:IS-KGBJS}.

\begin{figure}[t!]
\vspace{2.5cm}
  \begin{picture}(30,0)
    \put(10, -105){
      \includegraphics{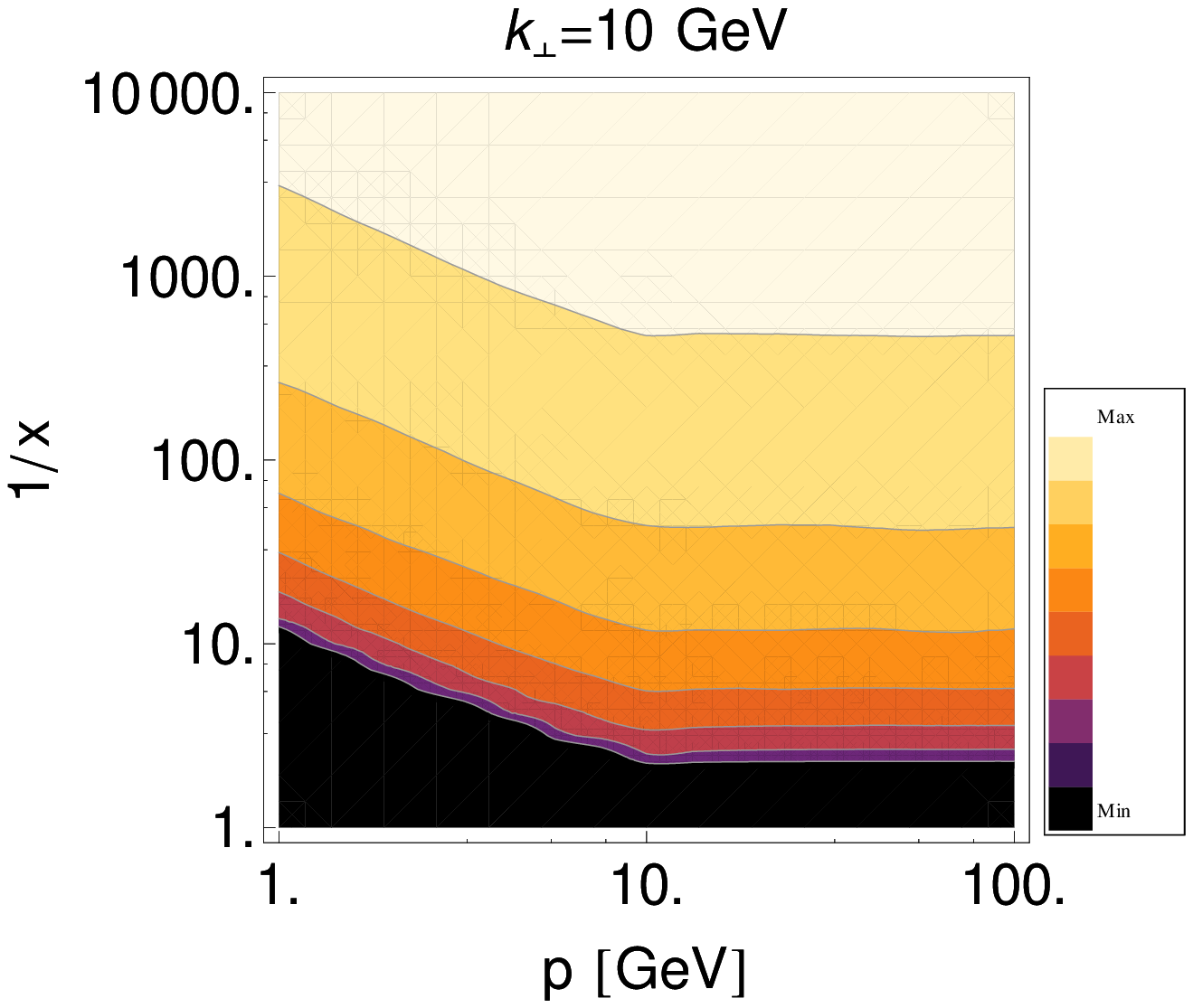}
    }
    \put(222, -105){
      \includegraphics{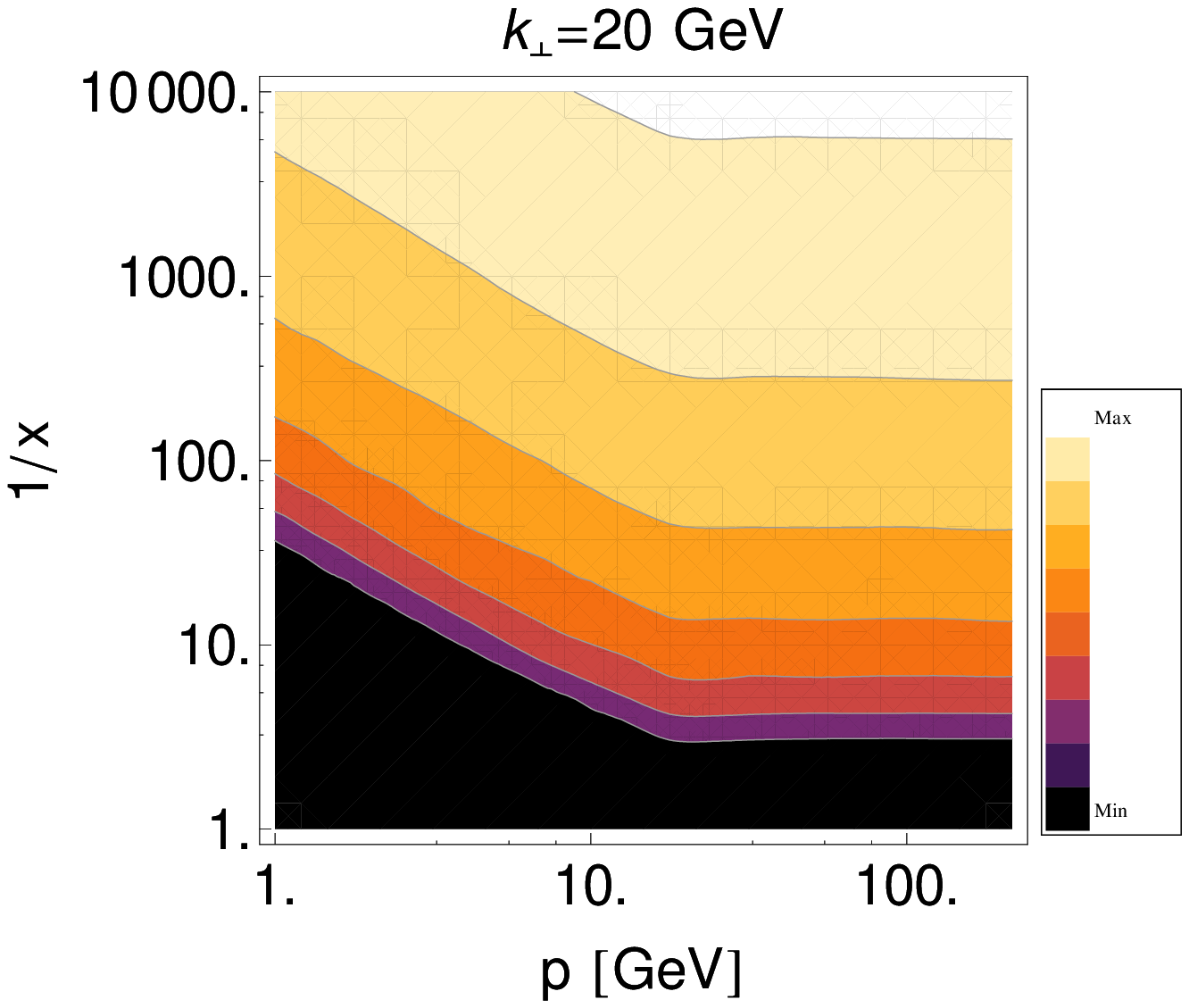}
      }
          
     \end{picture}
\vspace{3.6cm}
\caption{Relative ratio of CCFM~\eqref{eq:ccfm-or1} and KGBJS~\eqref{eq:IS-KGBJS} 
solutions. Distributions with definite $k_T$ for varying value of $p$.}
\label{fig:plots2D-1}
\end{figure}

\begin{figure}[t!]
\vspace{2.5cm}
  \begin{picture}(30,0)
    \put(10, -105){
      \includegraphics{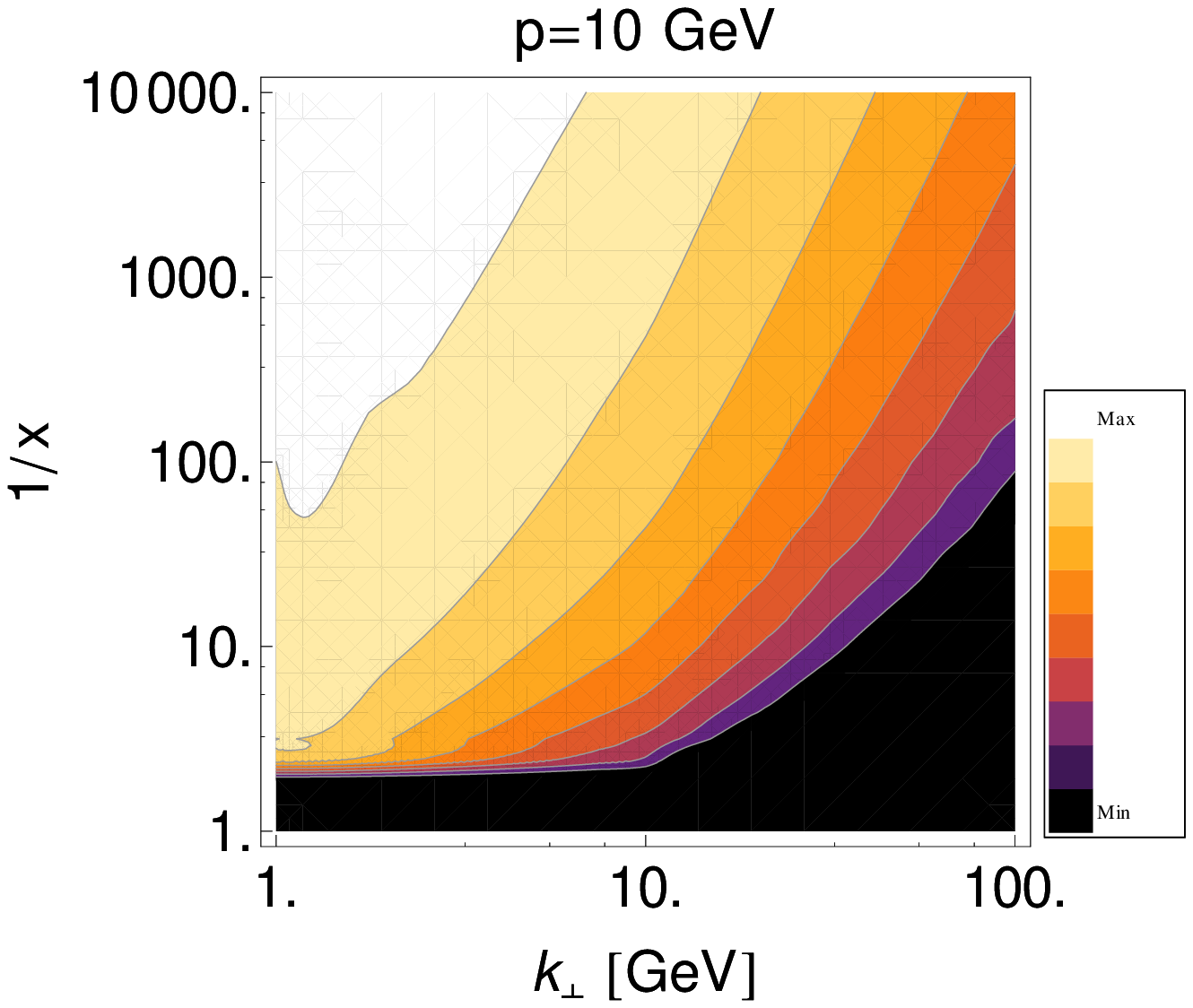}
    }    
    \put(222, -105){
      \includegraphics{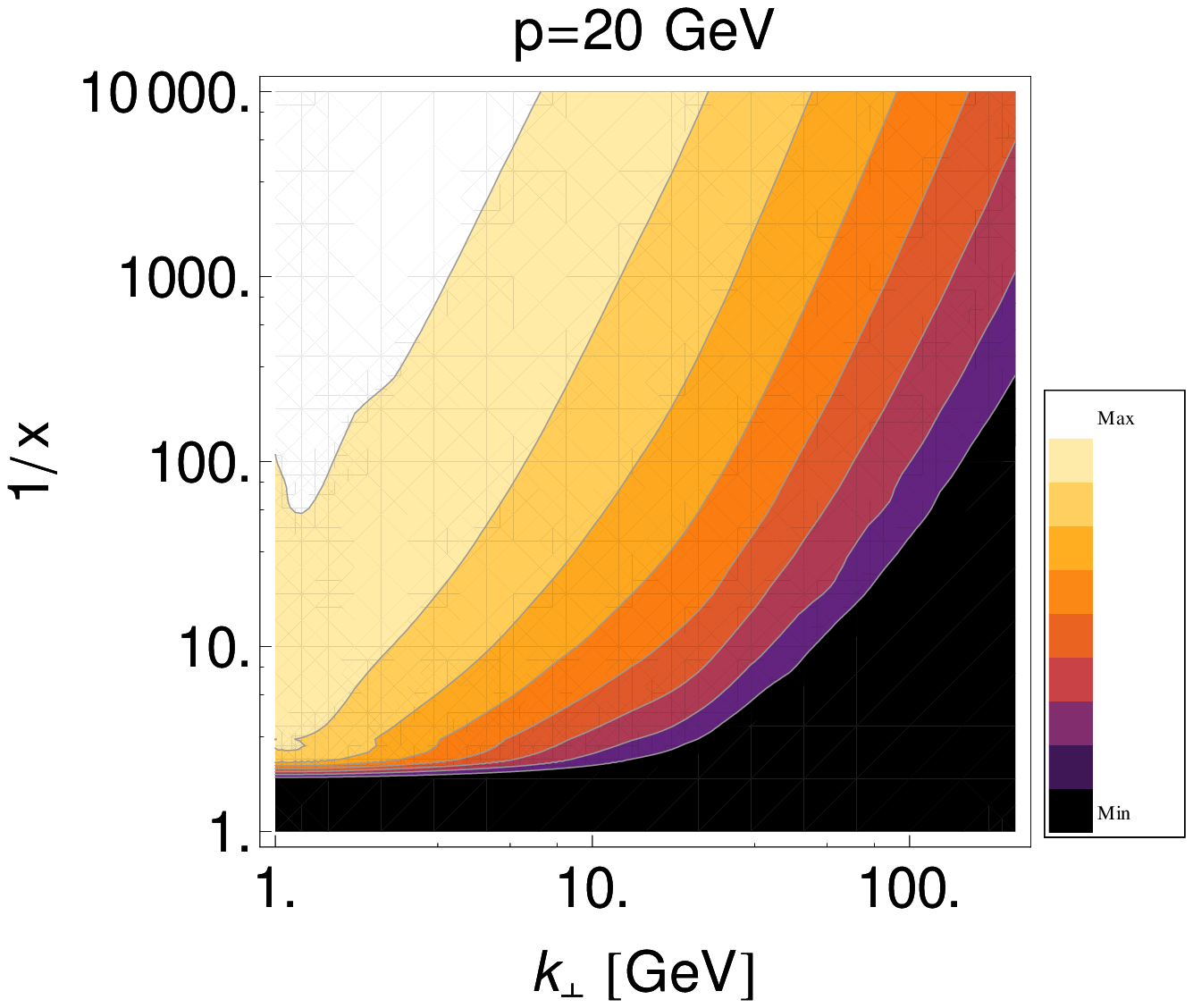}
    }
          
     \end{picture}
\vspace{3.6cm}
\caption{Relative ratio of CCFM~\eqref{eq:ccfm-or1} and KGBJS~\eqref{eq:IS-KGBJS} 
solutions. Distributions with definite $p$ for varying value of $k_T$.}
\label{fig:plots2D-2}
\end{figure}

By comparing the CCFM and KGBJS equations we see that the equations 
give quite similar distributions. This effect (for our choice of the starting 
scale $Q_0$) comes from the fact that the kinematical constraint suppresses the 
growth of the gluon so much that the non-linear effects enter only at very low $x$.
Observations made in previous paragraphs are confirmed in 2-dimensional plots
(Fig.~\ref{fig:plots2D-1}--\ref{fig:plots2D-2}), where we plot absolute 
relative difference of two amplitudes, solutions of the CCFM and the KGBJS 
equations, defined by the quantity
\begin{equation}
\beta\left(x,{k_T},p\right)=\frac{|{\mathcal A}_{CCFM}\left(x,{k_T},p\right)-
{\mathcal A}_{KGBJS}\left(x,{k_T},p\right)|}
{{\mathcal A}_{CCFM}\left(x,{k_T},p\right)}\;.
\end{equation}
The function $\beta\left(x,{k_T},p\right)$, introduced before in~\cite{Kutak:2013yga}, can be used to measure the strength of the non-linear effects and to define a 
saturation scale using the condition:
\begin{equation}
\beta\left(x,Q_s\left(x,p\right),p\right)=const.
\end{equation}
or 
$p$-related saturation scale:
\begin{equation}
\beta\left(x,k_T,P_s\right)=const.
\end{equation}
The conditions above can be seen as equipotential lines in the 2-dimensional plots in Fig.~\ref{fig:plots2D-1}--\ref{fig:plots2D-2} where different equipotential lines correspond to different constants on the right-hand side of the equation above.
The change in the slope of the $\beta\left(x,Q_s\left(x,p\right),p\right)$ at
around $k_T=p$ reported in~\cite{Kutak:2013yga}, 
apparent in Fig.~\ref{fig:plots2D-1}--\ref{fig:plots2D-2}, can be
understood in the context of the peak at $p\sim k_T$ 
(Fig.~\ref{fig:plots1D-1}--\ref{fig:plots1D-2}). For $k_T>p$ the
contribution of the integral on the right-hand side of the CCFM equation
decreases and the gluon density is dominated by the initial condition.

By comparing the plots Fig.~\ref{fig:plots2D-1}--\ref{fig:plots2D-2} to analogous plots in \cite{Kutak:2013yga} we see that their main features are very similar. We therefore conclude that the $low-x$ approximation of the KGBJS and CCFM equations taken  in \cite{Kutak:2013yga} does not, at least, modify the relative difference between linear and non-linear equation since the modifications of the kernel did not spoil the saturation pattern visible in the full equation. 
\subsection*{Dependence on the starting scale}
\begin{figure}[t!]
\vspace{2.5cm}
  \begin{picture}(30,0)
    \put(0, -105){
      \includegraphics{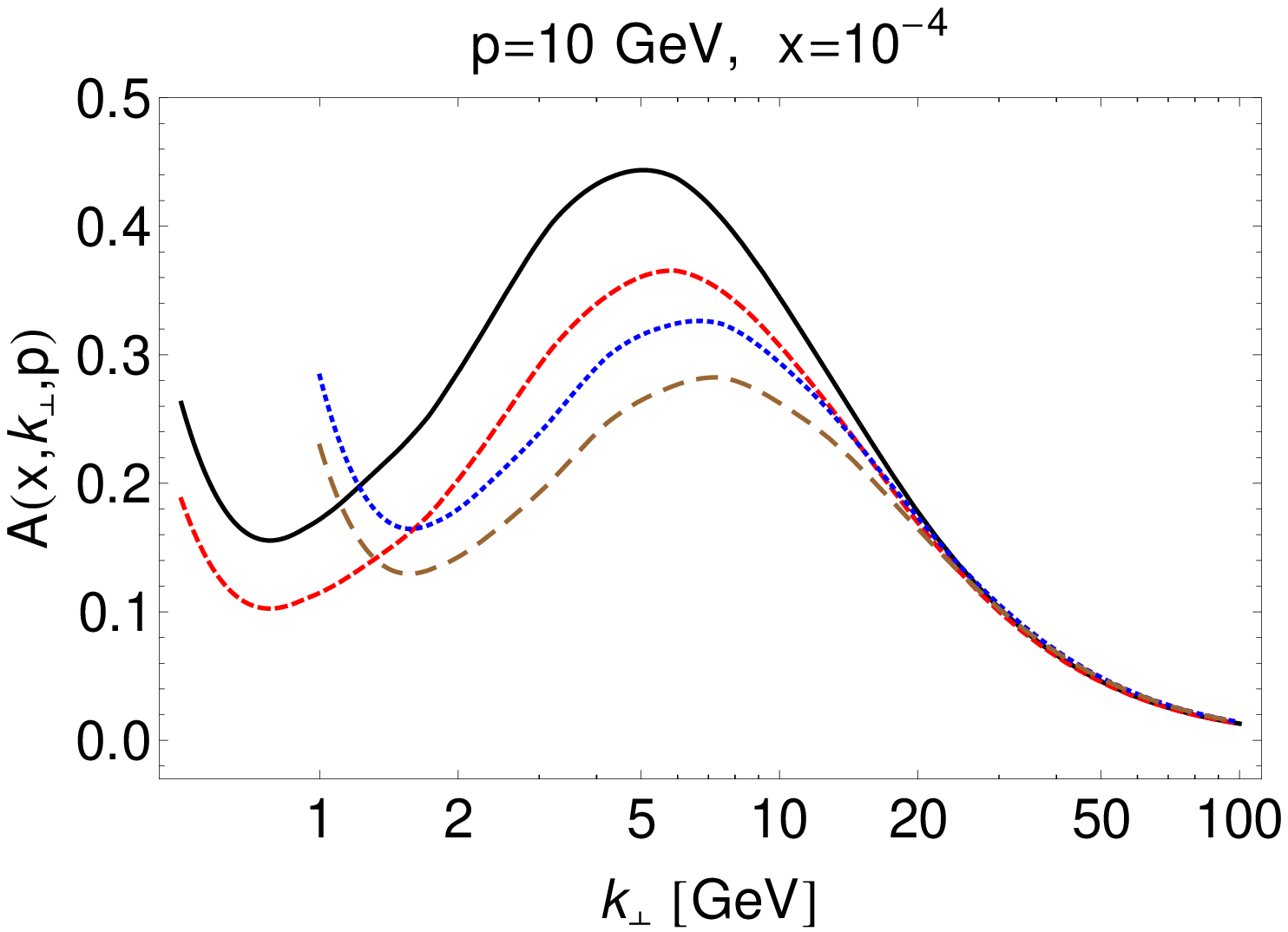}
    }    
    \put(222, -108){
      \includegraphics{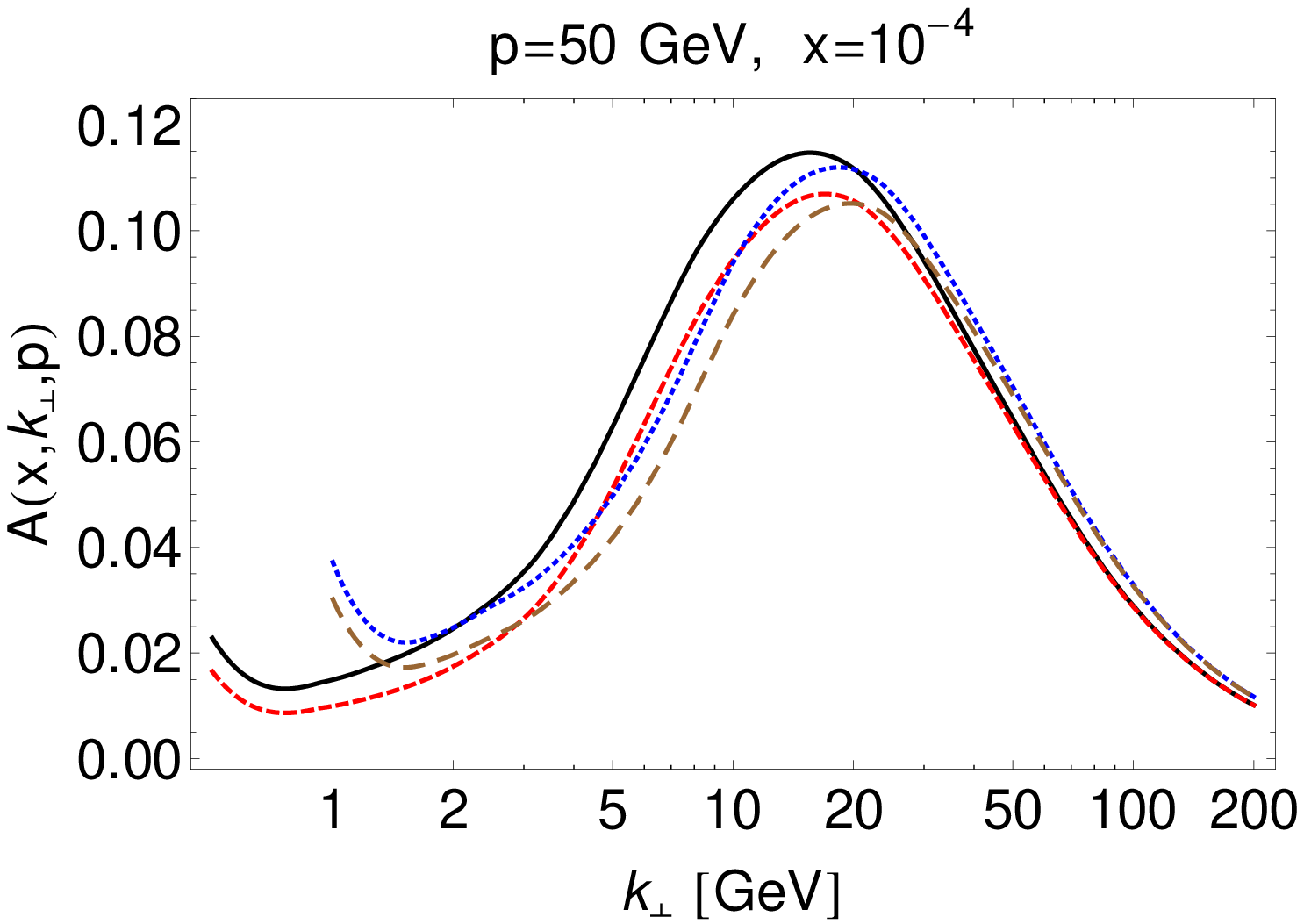}
    }
          
     \end{picture}
\vspace{3.6cm}
\caption{Solutions compared with $Q_0=0.5\;GeV$ and $Q_0=1\;GeV$. Black solid -
linear CCFM with $Q_0=0.5\;GeV$, red dashed - non-linear CCFM with $Q_0=0.5\;GeV$,
blue dotted - linear CCFM with $Q_0=1\;GeV$, brown dashed longer - linear CCFM with 
$Q_0=1\;GeV$.}
\label{fig:compared1}
\end{figure}
\begin{figure}[t]
\vspace{2.5cm}
  \begin{picture}(30,0)
    \put(0, -105){
      \includegraphics{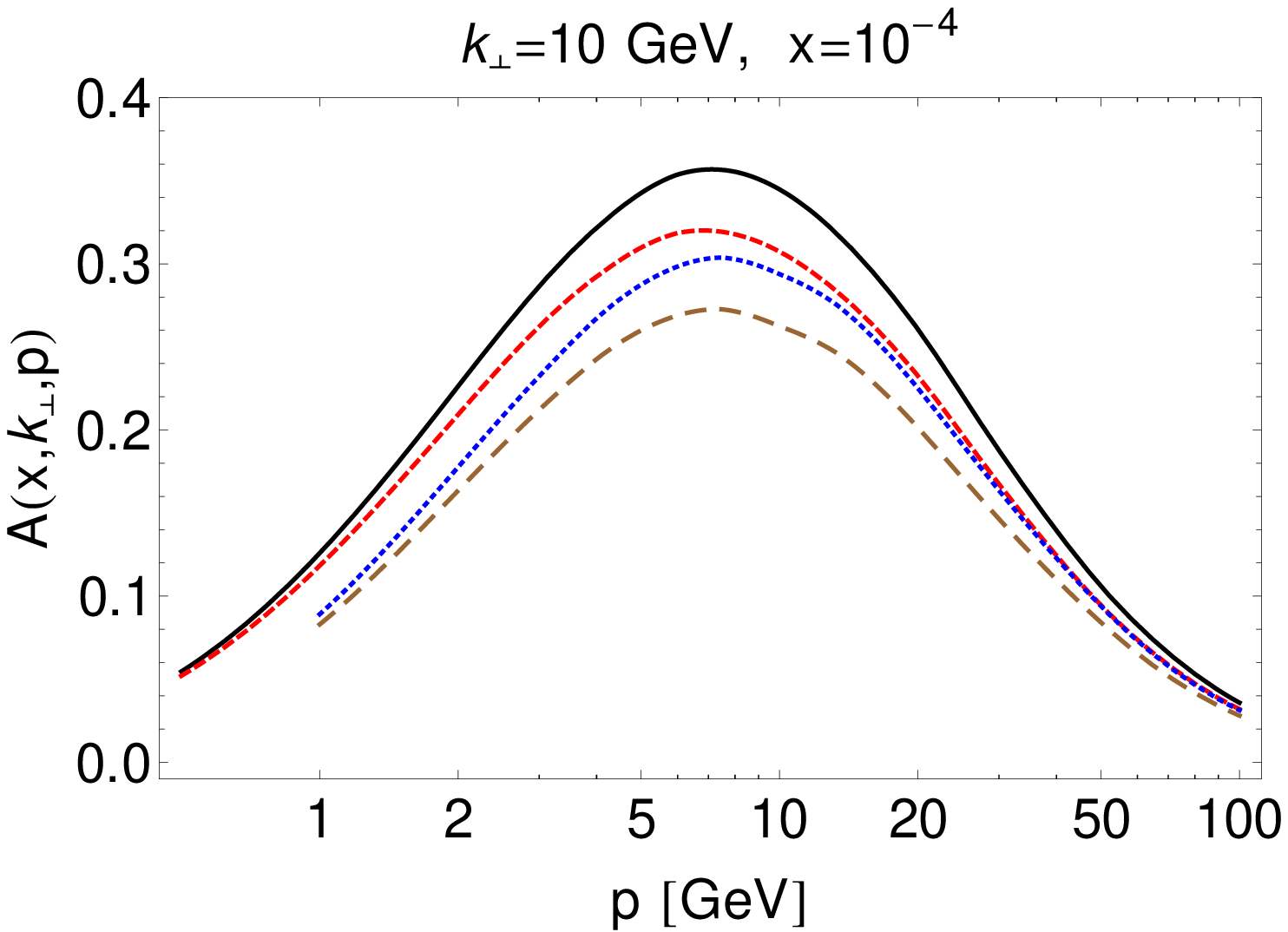}
    }    
    \put(222, -107){
      \includegraphics{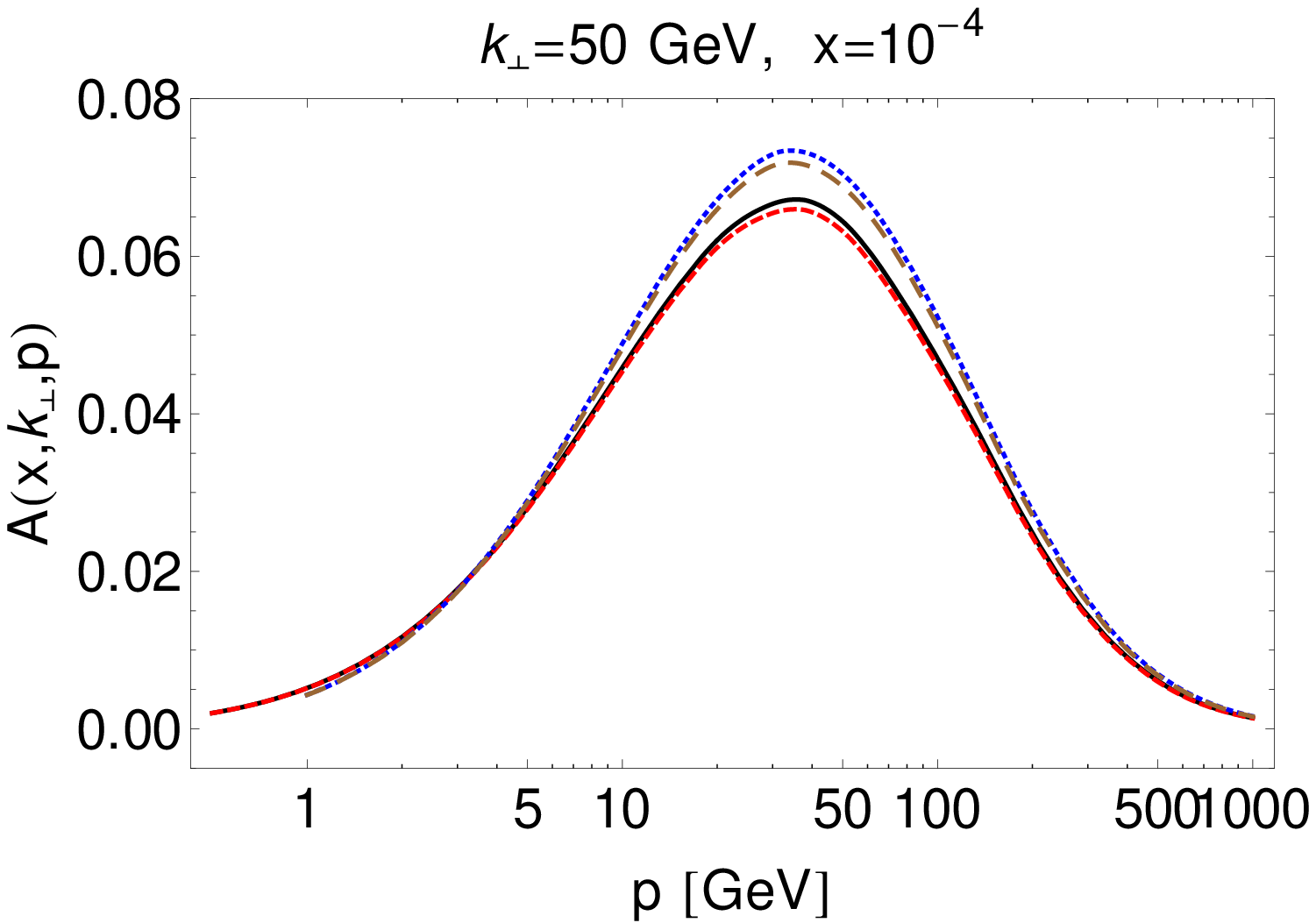}
    }
          
     \end{picture}
\vspace{3.6cm}
\caption{Solutions compared with $Q_0=0.5\;GeV$ and $Q_0=1\;GeV$. Black solid -
linear CCFM with $Q_0=0.5\;GeV$, red dashed - non-linear CCFM with $Q_0=0.5\;GeV$,
blue dotted - linear CCFM with $Q_0=1\;GeV$, brown dashed longer - linear CCFM with 
$Q_0=1\;GeV$.}
\label{fig:compared2}
\end{figure}
The starting scale $Q_0$ dependence is studied in Fig.~\ref{fig:compared1}-\ref{fig:compared2} using equations~\eqref{eq:ccfm-or1} and~\eqref{eq:IS-KGBJS}. 
We observe variation from few up to $25\%$ mostly in the local maximum of the $k_T$ and $p$ distributions. There is a peculiarity of the dependence on $Q_0$ of the $k_T$ distributions near the boundary $k_T=Q_0$ which can be explained by the dependence on the initial condition on the parameter $Q_0$. Near the point $k_T=Q_0$ the initial condition dominates and since it depends on the parameter the solutions also depend strongly on this parameter. We can tell, though, that for $k_T\gg p$ and $p\gg k_T$ the difference diminishes.

\section*{Conclusions and outlook}
We have solved linear and non-linear version of the CCFM equation with the full kernel, constant $\alpha_s$ and kinematical constraint included in two different ways.\\
Our numerical results make us conclude that the inclusion of the kinematical constraint in its full form into the
CCFM equation seems to affect largely the distribution of gluons. In particular inclusion of the kinematical constraint without omitting the $\theta$-function in the kernel of the equation causes strong suppression of the solution for all values of transversal momenta $k_T$. The effect is stronger than the implicit kinematical constraint included via modification of the non-Sudakov form-factor widely used in literature.
We expect that the suppression of the distribution will have significant effects on description of decorelations in azimuthal angle of forward-central jets \cite{Deak:2010gk,CMS:2014oma}. In particular we expect that the inclusion of the $\theta$-function is crucial for suppression of the cross section of emission of gluons in small angles since this configuration corresponds to large values of $k_T$. 
However, the conclusive comparison of the different approaches presented in this article can be done only after fitting the initial conditions  of the considered equations to data, this we, however, leave for future tasks.


\section*{Acknowledgements}
We would like to acknowledge useful discussions with Dawid Toton who participated in the project in the early stage.
The research of Krzysztof Kutak is supported by NCN grant DEC-2013/10/E/ST2/00656.
Michal Deak acknowledges support from Juan de la Cierva programme (JCI-2011-11382).


\begin{thebibliography}{99}

\bibitem{Gribov:1984tu} 
  L.~V.~Gribov, E.~M.~Levin and M.~G.~Ryskin,
  Phys.\ Rept.\  {\bf 100}, 1 (1983).
  
\bibitem{Catani:1990eg}
  S.~Catani, M.~Ciafaloni and F.~Hautmann,
  Nucl.\ Phys.\ B {\bf 366} (1991) 135.

\bibitem{Ciafaloni:1987ur}
  M.~Ciafaloni,
  Nucl.\ Phys.\ B {\bf 296} (1988) 49.
  
\bibitem{Catani:1989sg}
  S.~Catani, F.~Fiorani and G.~Marchesini,
  Nucl.\ Phys.\ B {\bf 336} (1990) 18.
  
\bibitem{Catani:1989yc}
  S.~Catani, F.~Fiorani and G.~Marchesini,
  Phys.\ Lett.\ B {\bf 234} (1990) 339.
\bibitem{Jung:2010si}
  H.~Jung, S.~Baranov, M.~Deak, A.~Grebenyuk, F.~Hautmann, M.~Hentschinski, A.~Knutsson and M.~Kramer {\it et al.},
  Eur.\ Phys.\ J.\ C {\bf 70} (2010) 1237
  [arXiv:1008.0152 [hep-ph]].
  
\bibitem{Hautmann:2013tba}
  F.~Hautmann and H.~Jung,
  Nucl.\ Phys.\ B {\bf 883} (2014) 1
  [arXiv:1312.7875 [hep-ph]].

\bibitem{Hautmann:2014uua}
  F.~Hautmann, H.~Jung and S.~T.~Monfared,
  Eur.\ Phys.\ J.\ C {\bf 74} (2014) 10,  3082
  [arXiv:1407.5935 [hep-ph]].

\bibitem{Kwiecinski:1996td}
  J.~Kwiecinski, A.~D.~Martin and P.~J.~Sutton,
  Z.\ Phys.\ C {\bf 71} (1996) 585
  [hep-ph/9602320].

\bibitem{Kutak:2011fu}
  K.~Kutak, K.~Golec-Biernat, S.~Jadach and M.~Skrzypek,
  JHEP {\bf 1202} (2012) 117
  [arXiv:1111.6928 [hep-ph]].

\bibitem{Kutak:2012yr}
  K.~Kutak,
  arXiv:1206.1223 [hep-ph].

\bibitem{Kutak:2012qk}
  K.~Kutak,
  JHEP {\bf 1212} (2012) 033
  [arXiv:1206.5757 [hep-ph]].

\bibitem{Deak:2009xt}
  M.~Deak, F.~Hautmann, H.~Jung and K.~Kutak,
  JHEP {\bf 0909} (2009) 121
  [arXiv:0908.0538 [hep-ph]].

\bibitem{Deak:2009ae}
  M.~Deak, F.~Hautmann, H.~Jung and K.~Kutak,
  arXiv:0908.1870 [hep-ph].
  
\bibitem{Deak:2010gk}
  M.~Deak, F.~Hautmann, H.~Jung and K.~Kutak,
  arXiv:1012.6037 [hep-ph].

\bibitem{Kutak:2012rf}
  K.~Kutak and S.~Sapeta,
  Phys.\ Rev.\ D {\bf 86} (2012) 094043
  [arXiv:1205.5035 [hep-ph]].

\bibitem{Chatrchyan:2012gwa}
  S.~Chatrchyan {\it et al.}  [CMS Collaboration],
  JHEP {\bf 1206} (2012) 036
  [arXiv:1202.0704 [hep-ex]].

\bibitem{vanHameren:2014ala}
  A.~van Hameren, P.~Kotko, K.~Kutak and S.~Sapeta,
  Phys.\ Lett.\ B {\bf 737} (2014) 335
  [arXiv:1404.6204 [hep-ph]].
\bibitem{vanHameren:2014lna}
  A.~van Hameren, P.~Kotko, K.~Kutak, C.~Marquet and S.~Sapeta,
  Phys.\ Rev.\ D {\bf 89} (2014) 9,  094014
  [arXiv:1402.5065 [hep-ph]].

%
%
%

\bibitem{Kutak:2014wga}
  K.~Kutak,
  Phys.\ Rev.\ D {\bf 91} (2015) 3,  034021
  [arXiv:1409.3822 [hep-ph]].

\bibitem{Deak:2012mx}
  M.~Deak,
  JHEP {\bf 1307} (2013) 087
  [arXiv:1209.6092 [hep-ph]].

\bibitem{Kutak:2013yga}
  K.~Kutak and D.~Toton,
  JHEP {\bf 1311} (2013) 082
  [arXiv:1306.3369 [hep-ph]].

\bibitem{Salam:1998cp}
  G.~P.~Salam,
  In *Brussels 1998, Deep inelastic scattering and QCD* 543-547
  [hep-ph/9805322].
\bibitem{Bottazzi:1998rs}
  G.~Bottazzi, G.~Marchesini, G.~P.~Salam and M.~Scorletti,
  JHEP {\bf 9812} (1998) 011
  [hep-ph/9810546].

\bibitem{Salam:1999ft}
  G.~P.~Salam,
  JHEP {\bf 9903} (1999) 009
  [hep-ph/9902324].
  
%
%
\bibitem{Avsar:2010ia}
  E.~Avsar and A.~M.~Stasto,
  JHEP {\bf 1006} (2010) 112
  [arXiv:1005.5153 [hep-ph]].
  

\bibitem{Bacchetta:2010hh}
  A.~Bacchetta, H.~Jung, A.~Knutsson, K.~Kutak and F.~Samson-Himmelstjerna,
  Eur.\ Phys.\ J.\ C {\bf 70} (2010) 503
  [arXiv:1001.4675 [hep-ph]].

\bibitem{Chachamis:2011rw}
  G.~Chachamis, M.~Deak, A.~S.~Vera and P.~Stephens,
  Nucl.\ Phys.\ B {\bf 849} (2011) 28
  [arXiv:1102.1890 [hep-ph]].

\bibitem{CMS:2014oma}
  CMS Collaboration [CMS Collaboration],
  CMS-PAS-FSQ-12-008.



\end{thebibliography}
\end{document}